\documentclass[acus]{pac99}
\usepackage{epsfig}

\newcommand{\ud}{\mathrm{d}}
\begin{document}
\title{NONLINEAR ACCELERATOR PROBLEMS VIA WAVELETS:\\
2. ORBITAL DYNAMICS IN GENERAL MULTIPOLAR FIELD}
\author{A.~Fedorova, M.~Zeitlin, IPME, RAS, St.~Petersburg, Russia
\thanks{e-mail: zeitlin@math.ipme.ru}
\thanks{http://www.ipme.ru/zeitlin.html;
        http://www.ipme.nw.ru/zeitlin.html}}
\maketitle
\begin{abstract}
In this series of eight papers  we
present the applications of methods from
wavelet analysis to polynomial approximations for
a number of accelerator physics problems.
In this part we consider orbital motion in transverse plane for a single
particle in a circular magnetic lattice in case when we take into account
multipolar expansion up to an arbitrary finite number. We reduce initial
dynamical problem to the finite number (equal to the number of n-poles) of standard
algebraical problem and represent all dynamical variables via an expansion in the
base of periodical wavelets.

\end{abstract}

\section{INTRODUCTION}
This is the second part of our eight presentations in which we consider
applications of methods from wavelet analysis to nonlinear accelerator
physics problems. This is a continuation of our results from [1]-[8],
which is based on our approach  to investigation
of nonlinear problems -- general, with additional structures (Hamiltonian,
symplectic or quasicomplex), chaotic, quasiclassical, quantum, which are
considered in the framework of local (nonlinear) Fourier analysis, or wavelet
analysis. Wavelet analysis is a relatively novel set of mathematical
methods, which gives us a possibility to work with well-localized bases in
functional spaces and with the general type of operators (differential,
integral, pseudodifferential) in such bases.
In this part we consider orbital motion in transverse plane for a single
particle in a circular magnetic lattice in case when we take into account
multipolar expansion up to an arbitrary finite number. We reduce initial
dynamical problem to the finite number (equal to the number of n-poles) of standard
algebraical problem and represent all dynamical variables as expansion in the
base of periodical wavelet functions. Our consideration is based on
generalization of variational wavelet approach from part 1. After introducing
 our starting points related to multiresolution in
section 3, we consider methods which allow us to construct wavelet representation for
solution in periodic case in section 4.

\section{Particle in the Multipolar Field}

The magnetic vector potential of a magnet with $2n$ poles in Cartesian
coordinates is
\begin{equation}
A=\sum_n K_nf_n(x,y),
\end{equation}
where $ f_n$ is a homogeneous function of  $x$ and $y$ of order $n$.
The real and imaginary parts of binomial expansion of
\begin{equation}
f_n(x,y)=(x+iy)^n
\end{equation}
correspond to regular and skew multipoles. The cases $n=2$ to $n=5$
correspond to low-order multipoles: quadrupole, sextupole, octupole, decapole.
Then we have in particular case the following equations of motion for
single particle in a circular magnetic lattice in the transverse plane
$(x,y)$ ([9] for designation):
\begin{eqnarray}
&&\frac{\ud^2x}{\ud s^2}+ \left(\frac{1}{\rho(s)^2}-k_1(s)\right)x=\nonumber\\
&&{\cal R}e\left[\sum_{n\geq 2}\frac{k_n(s)+ij_n(s)}{n!}\cdot(x+iy)^n\right],\\
&&\frac{\ud^2y}{\ud s^2}+k_1(s)y=\nonumber\\
&&-{\cal J}m\left[\sum_{n\geq}
\frac{k_n(s)+ij_n(s)}{n!}\cdot(x+iy)^n\right]\nonumber
\end{eqnarray}
and the corresponding Hamiltonian:
\begin{eqnarray}\label{eq:ham}
&&H(x,p_x,y,p_y,s)=\frac{p_x^2+p_y^2}{2}+\nonumber\\
&&\left(\frac{1}{\rho(s)^2}-k_1(s)\right)
\cdot\frac{x^2}{2}+k_1(s)\frac{y^2}{2}\\
&&-{\cal R}e\left[\sum_{n\geq 2}
\frac{k_n(s)+ij_n(s)}{(n+1)!}\cdot(x+iy)^{(n+1)}\right]\nonumber
\end{eqnarray}
Then we may take into account arbitrary but finite number in expansion
of RHS of Hamiltonian (\ref{eq:ham}) and
from our point of view the corresponding Hamiltonian equations of motions are
not more than nonlinear ordinary differential equations with polynomial
nonlinearities and variable coefficients.

\section{Wavelet Framework}
Our constructions are based on multiresolution approach. Because affine
group of translation and dilations is inside the approach, this
method resembles the action of a microscope. We have contribution to
final result from each scale of resolution from the whole
infinite scale of spaces. More exactly, the closed subspace
$V_j (j\in {\bf Z})$ corresponds to  level j of resolution, or to scale j.
We consider  a r-regular multiresolution analysis (MRA) of $L^2 ({\bf R}^n)$
(of course, we may consider any different functional space)
which is a sequence of increasing closed subspaces $V_j$:
\begin{equation}
...V_{-2}\subset V_{-1}\subset V_0\subset V_{1}\subset V_{2}\subset ...
\end{equation}
satisfying the following properties:
\begin{eqnarray}
&\displaystyle\bigcap_{j\in{\bf Z}}V_j=0,\quad
\overline{\displaystyle\bigcup_{j\in{\bf Z}}}V_j=L^2({\bf R}^n),\nonumber\\
&f(x)\in V_j <=> f(2x)\in V_{j+1}, \nonumber\\
&f(x)\in V_0 <=> f(x-k)\in V_0, \ \forall k\in {\bf Z}^n.
\end{eqnarray}
There exists a function $\varphi\in V_0$ such that
\{$\varphi_{0,k}(x)=$
$\varphi(x-k)$, $k\in{\bf Z}^n$\} forms a Riesz basis for $V_0$.
The function $\varphi$ is regular and localized:
$\varphi$ is $C^{r-1}$,\ $\varphi^{(r-1)}$ is almost
everywhere differentiable and for almost every $x\in {\bf R}^n$, for
every integer $\alpha\leq r$ and for all integer p there exists
constant $C_p$ such that
\begin{equation}
\mid\partial^\alpha \varphi(x)\mid \leq C_p(1+|x|)^{-p}
\end{equation}
Let
 $\varphi(x)$ be
a scaling function, $\psi(x)$ is a wavelet function and
$\varphi_i(x)=\varphi(x-i)$. Scaling relations that define
$\varphi,\psi$ are
\begin{eqnarray}
\varphi(x)&=&\sum\limits^{N-1}_{k=0}a_k\varphi(2x-k)=
\sum\limits^{N-1}_{k=0}a_k\varphi_k(2x),\\
\psi(x)&=&\sum\limits^{N-2}_{k=-1}(-1)^k a_{k+1}\varphi(2x+k).
\end{eqnarray}
Let  indices $\ell, j$
 represent translation and scaling, respectively and
\begin{equation}
\varphi_{jl}(x)=2^{j/2}\varphi(2^j x-\ell)
\end{equation}
then the set $\{\varphi_{j,k}\}, {k\in {\bf Z}^n}$ forms a Riesz basis for $V_j$.
The wavelet function $\psi $ is used to encode the details between
two successive levels of approximation.
Let $W_j$ be the orthonormal complement of $V_j$ with respect to $V_{j+1}$:
\begin{equation}
V_{j+1}=V_j\bigoplus W_j.
\end{equation}
Then just as $V_j$ is spanned by dilation and translations of the scaling function,
so are $W_j$ spanned by translations and dilation of the mother wavelet
$\psi_{jk}(x)$, where
\begin{equation}
\psi_{jk}(x)=2^{j/2}\psi(2^j x-k).
\end{equation}
All expansions which we used are based on the following properties:
\begin{eqnarray}
&&\{\psi_{jk}\}, \quad j,k\in {\bf Z}\quad
  \mbox{is a Hilbertian basis of } L^2({\bf R})\nonumber\\
&&\{\varphi_{jk}\}_{j\geq 0, k\in {\bf Z}} \quad\mbox{is an orthonormal
basis for} L^2({\bf R}),\nonumber\\
&& L^2({\bf R})=\overline{V_0\displaystyle\bigoplus^\infty_{j=0} W_j},\\
&& \mbox{or}\qquad
\{\varphi_{0,k},\psi_{j,k}\}_{j\geq 0,k\in {\bf Z}}\nonumber \\
&&\mbox{is
an orthonormal basis for}
 L^2({\bf R}).\nonumber
\end{eqnarray}
Fig.1 and Fig.2 give the representation of some function and
corresponding MRA on each level of resolution.

\begin{figure}[ht]
\centering
\epsfig{file=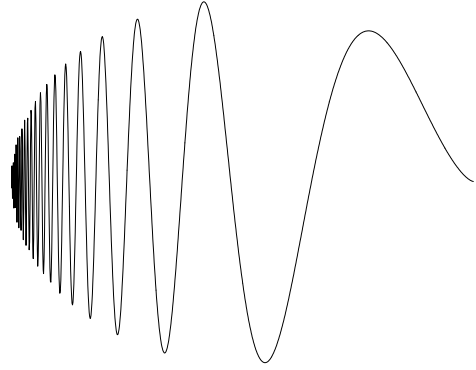, width=82.5mm, bb=0 200 599 590, clip}
\caption{Analyzed function.}
\end{figure}

\begin{figure}[ht]
\centering
\epsfig{file=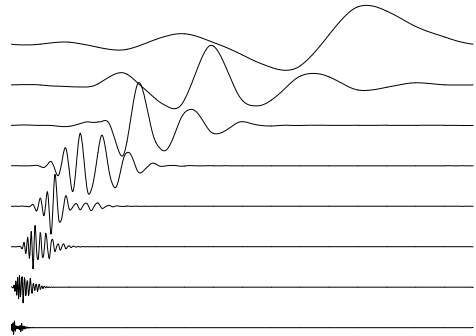, width=82.5mm, bb=0 200 599 590, clip}
\caption{MRA representation.}
\end{figure}

\section{VARIATIONAL WAVELET APPROACH\\ FOR PERIODIC TRAJECTORIES}
We start with extension of our approach from part 1 to the case
of periodic trajectories. The equations of motion corresponding
to Hamiltonian (4)  may also be formulated as a particular case of
the general system of
ordinary differential equations
$
{dx_i}/{dt}=f_i(x_j,t)$, $  (i,j=1,...,n)$, $0\leq t\leq 1$,
where $f_i$ are not more
than polynomial functions of dynamical variables $x_j$
and  have arbitrary dependence of time but with periodic boundary conditions.
According to our variational approach from part 1 we have the
solution in the following form
\begin{eqnarray}
x_i(t)=x_i(0)+\sum_k\lambda_i^k\varphi_k(t),\qquad x_i(0)=x_i(1),
\end{eqnarray}
where $\lambda_i^k$ are again the roots of reduced algebraical
systems of equations
with the same degree of nonlinearity and $\varphi_k(t)$
corresponds to useful type of wavelet bases (frames).
It should be noted that coefficients of reduced algebraical system
are the solutions of additional linear problem and
also
depend on particular type of wavelet construction and type of bases.
This linear problem is our second reduced algebraical problem. We need to find
in general situation objects
\begin{eqnarray}
\Lambda^{d_1 d_2 ...d_n}_{\ell_1 \ell_2 ...\ell_n}=
 \int\limits_{-\infty}^{\infty}\prod\varphi^{d_i}_{\ell_i}(x)\ud x,
\end{eqnarray}
but now in the case of periodic boundary conditions.
Now we consider the procedure of their
calculations in case of periodic boundary conditions
 in the base of periodic wavelet functions on
the interval [0,1] and corresponding expansion (14) inside our
variational approach. Periodization procedure
gives us
\begin{eqnarray}
\hat\varphi_{j,k}(x)&\equiv&\sum_{\ell\in Z}\varphi_{j,k}(x-\ell)\\
\hat\psi_{j,k}(x)&=&\sum_{\ell\in Z}\psi_{j,k}(x-\ell)\nonumber
\end{eqnarray}
So, $\hat\varphi, \hat\psi$ are periodic functions on the interval
 [0,1]. Because $\varphi_{j,k}=\varphi_{j,k'}$ if $k=k'\mathrm{mod}(2^j)$, we
may consider only $0\leq k\leq 2^j$ and as  consequence our
multiresolution has the form
$\displaystyle\bigcup_{j\geq 0} \hat V_j=L^2[0,1]$ with
$\hat V_j= \mathrm{span} \{\hat\varphi_{j,k}\}^{2j-1}_{k=0}$ [10].
Integration by parts and periodicity gives  useful relations between
objects (15) in particular quadratic case $(d=d_1+d_2)$:
$$
\Lambda^{d_1,d_2}_{k_1,k_2}=(-1)^{d_1}\Lambda^{0,d_2+d_1}_{k_1,k_2},\
\Lambda^{0,d}_{k_1,k_2}=\Lambda^{0,d}_{0,k_2-k_1}\equiv
\Lambda^d_{k_2-k_1}
$$
So, any 2-tuple can be represent by $\Lambda^d_k$.
Then our second additional linear problem is reduced to the eigenvalue
problem for
$\{\Lambda^d_k\}_{0\leq k\le 2^j}$ by creating a system of $2^j$
homogeneous relations in $\Lambda^d_k$ and inhomogeneous equations.
So, if we have dilation equation in the form
$\varphi(x)=\sqrt{2}\sum_{k\in Z}h_k\varphi(2x-k)$,
then we have the following homogeneous relations
\begin{equation}
\Lambda^d_k=2^d\sum_{m=0}^{N-1}\sum_{\ell=0}^{N-1}h_m h_\ell
\Lambda^d_{\ell+2k-m},
\end{equation}
or in such form
$A\lambda^d=2^d\lambda^d$, where $\lambda^d=\{\Lambda^d_k\}_
{0\leq k\le 2^j}$.
Inhomogeneous equations are:
\begin{equation}
\sum_{\ell}M_\ell^d\Lambda^d_\ell=d!2^{-j/2},
\end{equation}
 where objects
$M_\ell^d(|\ell|\leq N-2)$ can be computed by recursive procedure
\begin{eqnarray}
&&M_\ell^d=2^{-j(2d+1)/2}\tilde{M_\ell^d}, \\
&&\tilde{M_\ell^k}=
<x^k,\varphi_{0,\ell}>=\sum_{j=0}^k {k\choose j} n^{k-j}M_0^j,\quad
\tilde{M_0^\ell}=1.\nonumber
\end{eqnarray}
 So, we reduced our last problem to standard
linear algebraical problem. Then we use the same methods as in part 1.
As a result we obtained
for closed trajectories of orbital dynamics described by Hamiltonian (4)
the explicit time solution (14) in the base of periodized wavelets (16).

We are very grateful to M.~Cornacchia (SLAC), W.~Herrmannsfeldt (SLAC),
Mrs. J.~Kono (LBL) and
M.~Laraneta (UCLA) for their permanent encouragement

 \end{document}